\documentclass[12pt]{iopart}

\usepackage{iopams}  

\usepackage{cite}
\usepackage{bm}

\usepackage{color}

\begin{document}

\title{Duality in stochastic processes from the viewpoint of basis expansions}

\author{Jun Ohkubo$^{1,2}$, Yuuki Arai$^1$}

\address{$^1$ Graduate School of Science and Engineering, Saitama University, \\
255 Shimo-Okubo, Sakura-ku, Saitama, 338-8570, Japan}
\address{$^2$ JST, PRESTO, 4-1-8 Honcho, Kawaguchi, Saitama, 332-0012, Japan}
\ead{johkubo@mail.saitama-u.ac.jp}
\vspace{10pt}
\begin{indented}
\item[]
\end{indented}

\begin{abstract}
A new derivation method of duality relations in stochastic processes is proposed.
The current focus is on the duality between stochastic differential equations and birth-death processes.
Although previous derivation methods have been based on the viewpoint of time-evolution operators,
the current derivation is based on basis expansions.
In addition, only the tool needed for the derivation is the integration by parts,
which is rather simple and understandable.
The viewpoint of basis expansions enables us to obtain various dual stochastic processes.
As a demonstration, dual processes based on Taylor-type and Hermite polynomials are derived.
\end{abstract}

%
%
%
%
%

\section{Introduction}

Duality is a widely used concept in various research areas; 
for example, the Fourier transformation is an example of duality concepts,
which connects an original space and a frequency domain.
In stochastic processes, the duality concept has been used
in order to investigate interacting particle systems (for example, see \cite{Liggett_book}).
For example, a stochastic differential equation, which has 
a continuous-state space, is connected to a birth-death process with a discrete-state space.
Sometimes statistics for stochastic differential equations is evaluated
from the corresponding tractable birth-death processes,
so that the duality concept has been much investigated in various contexts
such as population genetics \cite{Shiga1986,Mohle1999,Carinci2015}, 
non-equilibrium heat-conduction problems \cite{Giardina2007,Carinci2013},
and simple exclusion processes \cite{Schutz1997,Imamura2011,Ohkubo2017}.

Recently, some mathematical discussions 
for the derivation of dual functions and dual processes
have been done; in \cite{Jansen2014}, recent developments have been reviewed.
Although there are discussions focusing on duality functions \cite{Redig2018},
the derivations have been mainly performed 
by using mathematical properties of time-evolution operators (generators);
the symmetry of the generators has been used to derive dual functions \cite{Giardina2009}.
In \cite{Ohkubo2010,Ohkubo2013}, 
the discussion based on the second-quantization method (the Doi-Peliti formalism) 
for the time-evolution operators has also been given.

In the present paper, a new viewpoint of basis expansions is proposed
in order to derive dual stochastic processes from stochastic differential equations.
By focusing on the basis, it is possible to view the duality concept more flexibly.
That is, we can use various types of basis expansions, and each expansion
has its own dual stochastic process;
consequently we can easily derive various types of dual stochastic processes.
In the present paper, as an example, 
we focus on a simple stochastic differential equation,
which is called as the stochastic logistic Ito equation in \cite{Doering2003}.
Using the stochastic logistic Ito equation, we demonstrate derivation of three different dual stochastic processes;
one is the conventional dual process; one is based on the Taylor-type expansion 
and it gives a slightly different process from the conventional one;
the final one is based on the Hermite polynomials 
and a completely different dual process is derived.
Here, we have a comment for the usage of the orthogonal polynomials.
As shown later, various orthogonal polynomials could be used to derive dual processes.
Although there are a few works about the connections
between the duality and orthogonality relations \cite{Franceschini2017,Groenevelt2019},
the proposition in the present paper is different from these works;
the usage of the orthogonal polynomials as the basis expansions is the main theme here.

The structure of the present paper is as follows.
In section 2, the brief review of the conventional duality concept is given.
Section 3 focuses on the mathematical structure of duality concept;
in order to understand the duality, it is enough to use the integration by parts
and basis expansions.
In section 4, the derivations of dual stochastic processes based on the basis expansions
are demonstrated using the concrete example.

\section{Brief summary of duality concept}

\subsection{Definition of the duality}

In the present paper, we focus on the duality relation between stochastic differential equations
and birth-death processes.
The stochastic differential equation has
a continuous-state and continuous-time; 
the state vector at time $t$ is given by $\bm{x}_t \in \mathbb{R}^{D_1}$, 
where $D_1$ is the dimension of the state vector.
The birth-death process, whose state vector at time $t$ is  $\bm{n}_t \in \mathbb{N}^{D_2}$, 
has a discrete-state and continuous-time.
Note that these two processes do not need to have the same dimensions.

The process $(\bm{x}_t)$ is said to be \textit{dual} to $(\bm{n}_t)$
with respect to a duality function $D: \mathbb{R}^{D_1} \times \mathbb{N}^{D_2} \to \mathbb{R}$
if for all $(\bm{x}_t)$, $(\bm{n}_t)$ 
and $t \ge 0$ we have
\begin{eqnarray}
\mathbb{E}_{\bm{n}_t} \left[ D(\bm{x}_0,\bm{n}_t)\right]
= \mathbb{E}_{\bm{x}_t} \left[ D(\bm{x}_t,\bm{n}_0)\right],
\label{eq_basic_duality}
\end{eqnarray}
where $\mathbb{E}_{\bm{x}_t}$ and $\mathbb{E}_{\bm{n}_t}$ are the expectations in the processes
$(\bm{x}_t)$ starting from $\bm{x}_0$
and
$(\bm{n}_t)$ starting from $\bm{n}_0$,
respectively.

\subsection{Problem settings}

Here, we focus on the stochastic logistic Ito equation in \cite{Doering2003} as an example.
As denoted in \cite{Doering2003}, the stochastic logistic Ito equation 
is related to the stochastic Fisher and Kolmogorov-Petrovsky-Piscounov (sFKPP) equation, 
which plays an important role in the study of the
front-propagation problems \cite{Brunet1997,Pechenik1999,Panja2004,Brunet2006}
and QCD context \cite{Munier2006}.
The spatial part in the sFKPP equation is neglected in the stochastic logistic Ito equation,
and the dual stochastic process has also been known.
Here, we employ the notation in \cite{Ohkubo2013}.

The target stochastic differential equation is expressed as 
the following Ito-type stochastic differential equation 
\begin{eqnarray}
\rmd x = - \gamma x(1-x) \rmd t + \sigma \sqrt{x(1-x)} \rmd W(t)
\label{eq_sFKPP}
\end{eqnarray}
for $0 \le x \le 1$,
where $\gamma$ and $\sigma$ are parameters, and $W(t)$ expresses a Wiener process.
Note that a variable transformation $u = 1 - x$ recovers the previous discussions in \cite{Doering2003}.

The corresponding partial differential equation (the Fokker-Planck equation) is as follows:
\begin{eqnarray}
\frac{\partial}{\partial t} p(x,t)
= - \frac{\partial}{\partial x} 
\left[ - \gamma x(1-x) p(x,t)\right]
+ \frac{1}{2} \frac{\partial^2}{\partial x^2}
\left[ \sigma^2 x(1-x) p(x,t) \right],
\label{eq_Fokker_Planck}
\end{eqnarray}
where $p(x,t)$ is the probability density function at time $t$.

\subsection{Conventional duality}

There are some discussions for the derivation of the dual stochastic process;
for example, see \cite{Giardina2009}.
If focusing on the stochastic differential equations,
the so-called Doi-Peliti method \cite{Doi1976,Doi1976a,Peliti1985} 
achieves a systematic derivation \cite{Ohkubo2013};
the Doi-Peliti method is based on the second-quantization in quantum mechanics,
and creation and annihilation operators play important roles in the derivation.
Note that the details are not written here because the new derivation method proposed 
in the present paper does not need the knowledge of the Doi-Peliti method.
We give only the final consequences of the discussion;
finally, for the system in \eref{eq_sFKPP},
we obtain the following dual stochastic process
denoted as a birth-coagulation process for particles $A$:
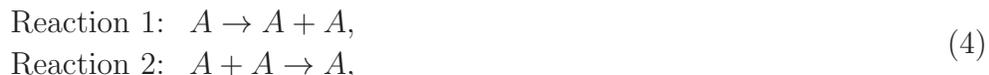
\begin{eqnarray}
\begin{array}{l}
\textrm{Reaction 1: } \,\, A \to A + A,\\
\textrm{Reaction 2: } \,\, A + A \to A,
\end{array}
\label{eq_naive_chemical_reactions}
\end{eqnarray}
i.e.,
\begin{eqnarray}
\begin{array}{ll}
n \to n+1 & \textrm{at rate} \,\, \gamma n, \\
n \to n-1 & \textrm{at rate} \,\, \sigma^2 n (n-1) / 2,
\end{array}
\label{eq_naive_birth_death}
\end{eqnarray}
where $n$ is the number of particles $A$.
The master equation for the birth-coagulation process
is written as follows:
\begin{eqnarray}
\frac{\rmd}{\rmd t} P(n,t) 
= &\gamma(n-1)P(n-1,t) - \gamma n P(n,t) \nonumber \\
&+ \frac{\sigma^2}{2} n(n+1) P(n+1,t) - \frac{\sigma^2}{2} (n-1)n P(n,t),
\label{eq_naive_master_equation}
\end{eqnarray}
where $P(n,t)$ is the probability distribution 
for the state with $n$ particles at time $t$.

Through the dual function $D(x,n) = x^n$,
the original stochastic differential equation in \eref{eq_sFKPP}
is connected with the birth-coagulation process in \eref{eq_naive_master_equation} as follows:
\begin{eqnarray}
\mathbb{E}_{x_t}\left[ x^m \right]
= \int_0^1 \rmd x \, p(x,t) x^m 
= \sum_{n=0}^\infty P(n,t) x_0^{n},
\end{eqnarray}
where the initial condition for the birth-coagulation process should be
\begin{eqnarray}
P(m,t=0) = 1 \quad \textrm{and} \quad
P(n,t=0) = 0 \quad \textrm{for $n \neq m$},
\end{eqnarray}
and the initial condition for the stochastic differential equation
should be 
\begin{eqnarray}
p(x,t=0) = \delta(x-x_0),
\end{eqnarray}
where $x_0$ is the initial position, and $\delta(x)$ is the Dirac delta function.
That is, once we solve the dual stochastic process in \eref{eq_naive_master_equation},
we can immediately obtain the $m$-th moment in the stochastic differential equation
\textit{for arbitrary initial conditions};
it is not necessary to perform simulations with different initial conditions
for the stochastic differential equation.
This property has been exploited in statistical physics \cite{Giardina2009},
and there is also a numerical application of this property in nonlinear Kalman filtering \cite{Ohkubo2015}.

In \cite{Ohkubo2013}, further discussions
for slightly different stochastic differential equations were given;
some extensions of the duality concept are needed
in order to recover probabilistic property 
and to deal with negative transition rates; for details, see \cite{Ohkubo2013}.
Although these extensions have been done via some techniques 
for the time-evolution operators written in the creation and annihilation operators,
in the following discussions in the present paper,
simple ways for the same extensions will be shown.

\section{Rewriting the duality concept from the viewpoint of basis expansions}

This section gives one of the main contributions of the present paper.
From the viewpoint of basis expansions, it is straightforward to understand
the derivation of the duality relations
between stochastic differential equations and birth-death processes.
For readability, we here restrict our discussion to one variable cases.
It is straightforward to extend the discussion 
to multivariate cases.

Here, as denoted above, we interest in the $m$-th moment
of the stochastic differential equation.
Suppose that $p(x,t)$ is the probability density distribution 
for the stochastic differential equation
and the time-evolution for the corresponding Fokker-Planck equation 
is given as a time evolution operator $\mathcal{L}$,
a formal solution of the Fokker-Planck equation is written as
\begin{eqnarray}
p(x,t) = \rme^{\mathcal{L} t} p(x,t=0) = \rme^{\mathcal{L} t} \delta(x-x_0),
\end{eqnarray}
where we suppose that the initial position
of the stochastic differential equation is $x = x_0$.
Hence, the calculation of the $m$-th moment is rewritten as follows:
\begin{eqnarray}
\mathbb{E}\left[ x^m \right] 
&= \int_{-\infty}^{\infty} x^m p(x,t) \rmd x \nonumber \\
&= \int_{-\infty}^{\infty} x^m \left( \rme^{\mathcal{L} t} \delta(x-x_0)\right) \rmd x \nonumber \\
&= \int_{-\infty}^{\infty} \left( \rme^{\mathcal{L}^\dagger t} x^m \right)  \delta(x-x_0) \rmd x \nonumber \\
&= \int_{-\infty}^{\infty} \widetilde{p}(x,t) \delta(x-x_0) \rmd x \nonumber \\
&= \widetilde{p}(x_0,t),
\label{eq_derivation_expectation}
\end{eqnarray}
where $\mathcal{L}^\dagger$ is the adjoint operator of $\mathcal{L}$,
and $\widetilde{p}(x,t)$ is obtained as a result of 
the time-evolution using the adjoint operator $\mathcal{L}^\dagger$.
Note that $\widetilde{p}(x,t)$ is not a probability density distribution in general.

The adjoint operator is easily derived from the integration by parts;
for a pedagogical purpose, we here describe more details.
Firstly, consider the following conventional stochastic differential equation:
\begin{eqnarray}
\rmd x = A(x,t) \rmd t + B(x,t) \rmd W(t).
\end{eqnarray}
Then, the time-evolution operator for the corresponding Fokker-Planck equation
is given as follows \cite{Gardiner_book}:
\begin{eqnarray}
\mathcal{L} = - \frac{\partial}{\partial x} D^{(1)}(x,t) 
+ \frac{\partial^2}{\partial x^2} D^{(2)}(x,t),
\label{eq_derivation_L}
\end{eqnarray}
where 
\begin{eqnarray}
D^{(1)}(x,t) = A(x,t), \quad D^{(2)}(x,t) = \frac{1}{2} \left( B(x,t) \right)^2.
\end{eqnarray}
Here focusing on the first term in \eref{eq_derivation_L}, we have
\begin{eqnarray}
\fl
-\int_{-\infty}^{\infty} \rmd x \, \widetilde{p}(x,t)
\frac{\partial}{\partial x} \left( D^{(1)}(x,t) p(x,t) \right)  \nonumber \\
\fl \qquad
= - \left[ \widetilde{p}(x,t) D^{(1)}(x,t) p(x,t)  \right]_{-\infty}^{\infty}
+ \int_{-\infty}^{\infty} \rmd x \, \left( \frac{\partial}{\partial x}  \widetilde{p}(x,t) \right) 
D^{(1)}(x,t) p(x,t) \nonumber \\
\fl \qquad
= \int_{-\infty}^{\infty} \rmd x \, 
\left\{ D^{(1)}(x,t) \left( \frac{\partial}{\partial x}  \widetilde{p}(x,t) \right) \right\}
p(x,t),
\end{eqnarray}
where we used the integration by parts
and the fact that the probability density function $p(x,t)$ goes to $0$ when $x \to \pm \infty$.
After employing the same calculation for the second term in \eref{eq_derivation_L}, 
we have the adjoint time-evolution operator $\mathcal{L}^\dagger$ as follows:
\begin{eqnarray}
\mathcal{L}^\dagger = 
D^{(1)}(x,t) \frac{\partial}{\partial x} + D^{(2)}(x,t) \frac{\partial^2}{\partial x^2}.
\end{eqnarray}

From \eref{eq_derivation_expectation}, 
it is clear that 
it is enough to perform a time-evolution with $\mathcal{L}^\dagger$ and the initial condition $x^m$
instead of the time-evolutions for the original system with various initial conditions 
($\delta(x-x_0)$, i.e., the particle starts from the position $x_0$);
once we obtain $\widetilde{p}(x_0,t)$, 
the $m$-th moment of the original stochastic differential equation 
for \textit{various} initial conditions is evaluated.
However, note that $\widetilde{p}(x,t)$ does not correspond to a birth-death process;
$x$ is a continuous variable.
In order to recover the discrete characteristics of the dual stochastic process,
we need a basis expansion as follows:
\begin{eqnarray}
\widetilde{p}(x,t) = \sum_{n=0}^\infty \widetilde{P}(n,t) \phi_n(x),
\end{eqnarray}
where $\{\phi_n(x)\}$ are basis functions
and $\{\widetilde{P}(n,t)\}$ are expansion coefficients.
By using an adequate basis $\{\phi_n(x)\}$,
we can obtain the time-evolution equations
for the coefficients $\widetilde{P}(n,t)$ from the adjoint time-evolution operator $\mathcal{L}^\dagger$.

Of course, in general, $\widetilde{P}(n,t)$ is not a probability distribution,
and some techniques are needed to interpret 
the time-evolution equations for $\widetilde{P}(n,t)$ 
as a dual stochastic process (birth-death processes) with discrete characteristics.
We will demonstrate the techniques for the interpretation in the next section.

\section{Demonstration and some techniques to recover probabilistic properties}

\subsection{Restatement of the problem}

For readers convenient, here a concise statement of the problem is denoted again.

The main aim is to evaluate the $m$-th moment $\mathbb{E}\left[ x^m \right]$
of the stochastic differential equation in \eref{eq_sFKPP} at time $t$.
In order to perform the evaluation, 
it is enough to solve the following partial differential equation:
\begin{eqnarray}
\frac{\partial}{\partial t} \widetilde{p}(x,t)
= - \gamma x(1-x) \frac{\partial}{\partial x} \widetilde{p}(x,t)
+ \frac{\sigma^2}{2} x(1-x) \frac{\partial^2}{\partial x^2} \widetilde{p}(x,t),
\label{eq_ex_adjoint}
\end{eqnarray}
where the initial condition for $\widetilde{p}(x,t)$ should be
\begin{eqnarray}
\widetilde{p}(x,t=0) = x^m.
\label{eq_ex_initial}
\end{eqnarray}
Solving \eref{eq_ex_adjoint},
the $m$-th moment with an arbitrary initial condition $x = x_0$
is immediately given as $\widetilde{p}(x_0,t)$.

\subsection{Conventional duality}

As reviewed in section 2,
the conventional viewpoint of the time-evolution operator 
gives the dual stochastic process 
(the birth-death process or the birth-coagulation process) 
in \eref{eq_naive_birth_death}.
We here show that a simple power-expansion 
recovers the dual stochastic process immediately.

Employing the basis function
\begin{eqnarray}
\phi_n(x) = x^n,
\end{eqnarray}
we obtain
\begin{eqnarray}
\widetilde{p}(x,t) = \sum_{n=0}^\infty P(n,t) x^n.
\label{eq_ex_power}
\end{eqnarray}
Inserting \eref{eq_ex_power} into \eref{eq_ex_adjoint}, we have
\begin{eqnarray}
\fl
\frac{\rmd}{\rmd t}
\sum_{n=0}^\infty P(n,t) x^n
&= -\gamma x (1-x) \frac{\rmd}{\rmd x} \sum_{n=0} P(n,t) x^n
+ \frac{\sigma^2}{2} x(1-x) \frac{\rmd^2}{\rmd x^2} \sum_{n=0} P(n,t) x^n
\nonumber \\
\fl
&= \gamma \sum_{n=0}^\infty (n-1)P(n-1,t) x^n
- \gamma \sum_{n=0}^\infty n P(n,t) x^n \nonumber \\
\fl
&\quad + \frac{\sigma^2}{2} \sum_{n=0} (n+1)n P(n+1,t) x^n
- \frac{\sigma^2}{2} \sum_{n=0} n(n-1) P(n,t) x^n,
\end{eqnarray}
and hence, 
by comparing the coefficients with the same degree in $x^n$,
we recover the time-evolution equation in \eref{eq_naive_master_equation} for $\{P(n,t)\}$.

\subsection{Taylor-type basis functions}

Here, the following Taylor-type basis expansion is employed:
\begin{eqnarray}
\phi_n(x) = \frac{x^n}{n!},
\end{eqnarray}
so that,
\begin{eqnarray}
\widetilde{p}(x,t) = \sum_{n=0}^\infty  P_{\mathrm{T}}(n,t) \frac{x^n}{n!},
\label{eq_ex_Taylor_expansion}
\end{eqnarray}
where $\{P_{\mathrm{T}}(n,t)\}$ are coefficients for the Taylor-type case.

Using the same discussion with section 4.2,
we have the following time-evolution equation for the coefficients $\{P_{\mathrm{T}}(n,t)\}$:
\begin{eqnarray}
\frac{\rmd}{\rmd t} P_{\mathrm{T}}(n,t) 
= &\gamma n(n-1) P_{\mathrm{T}}(n-1,t) - \gamma n P_{\mathrm{T}}(n,t) \nonumber \\
&+ \frac{\sigma^2}{2} n P_{\mathrm{T}}(n+1,t) - \frac{\sigma^2}{2} (n-1)n P_{\mathrm{T}}(n,t).
\label{eq_dual_basic_time_evolution}
\end{eqnarray}
Although this equation does not satisfy the probability conservation law,
as discussed in \cite{Ohkubo2013}, it is possible to recover the probabilistic characteristics
and we can use Monte Carlo simulations (for example, the Gillespie algorithm \cite{Gillespie1977})
in order to obtain the coefficients $\{P_{\mathrm{T}}(n,t)\}$.
We briefly review the procedures to use the Monte Carlo simulations.
The basic principle for the rewriting of the time-evolution equation is as follows:
\begin{enumerate}
\item Separate the terms into the following two parts:
\begin{enumerate}
\item[(A)] Terms in which the state is changed;
\item[(B)] Terms in which the state is not changed.
\end{enumerate}
\item Make the part (A) satisfy the probability conservation law
by subtracting some terms.
For the compensation,
the corresponding terms are added to the part (B).
\end{enumerate}
Using \eref{eq_dual_basic_time_evolution}, we demonstrate the procedures.
Note that the first and third terms in the r.h.s
in \eref{eq_dual_basic_time_evolution} have $P_{\mathrm{T}}(n-1,t)$
and $P_{\mathrm{T}}(n+1,t)$,
and hence the states are changed as $n \to n-1$ and $n \to n+1$ respectively.
In contrast, the second and fourth terms only have $P_{\mathrm{T}}(n,t)$,
which means that these terms do not change the states.
Hence, we have
\begin{eqnarray}
\frac{\rmd}{\rmd t} P_{\mathrm{T}}(n,t) 
= &\left\{ \gamma n(n-1) P_{\mathrm{T}}(n-1,t) 
+ \frac{\sigma^2}{2} n P_{\mathrm{T}}(n+1,t) \right\} \nonumber \\
& + \left\{ - \gamma n P_{\mathrm{T}}(n,t) 
- \frac{\sigma^2}{2} (n-1)n P_{\mathrm{T}}(n,t) \right\}.
\end{eqnarray}
Focusing on terms in the first curly bracket,
we need to subtract two terms in order to satisfy the probability conservation law for this part,
and the subtracted terms are added to the second curly bracket;
\begin{eqnarray}
\frac{\rmd}{\rmd t} P_{\mathrm{T}}(n,t) 
=& \Big\{ \gamma n(n-1)  P_{\mathrm{T}}(n-1,t)
- \gamma (n-1)(n-2) P_{\mathrm{T}}(n,t) \nonumber \\
&+ \frac{\sigma^2}{2} n P_{\mathrm{T}}(n+1,t)
- \frac{\sigma^2}{2} (n-1) P_{\mathrm{T}}(n,t) \Big\} \nonumber \\
&+ \Big\{ - \gamma n P_{\mathrm{T}}(n,t) 
- \frac{\sigma^2}{2} n(n-1) P_{\mathrm{T}}(n,t) \nonumber \\
&+ \gamma (n-1)(n-2) P_{\mathrm{T}}(n,t) 
+ \frac{\sigma^2}{2} (n-1) P_{\mathrm{T}}(n,t) \Big\}.
\label{eq_ex_Taylor_master_2}
\end{eqnarray}
The first term in the r.h.s. in \eref{eq_ex_Taylor_master_2} satisfies
the probability conservation law,
and it corresponds to the following birth-death process:
\begin{eqnarray}
\begin{array}{ll}
n \to n+1 & \textrm{at rate} \,\, \gamma (n-1)(n-2), \\
n \to n-1 & \textrm{at rate} \,\, \sigma^2 (n-1)/2.
\end{array}
\label{eq_ex_Taylor_birth_death}
\end{eqnarray}
Of course, the second term in the r.h.s. in \eref{eq_ex_Taylor_master_2} 
should be dealt with adequately;
making $N$ sample paths via the Monte Carlo simulations 
and denoting $i$-th path as $\{n^{(i)}_{t}\}$,
we have
\begin{eqnarray}
P_{\mathrm{T}}(n,t) = 
\frac{w_\mathrm{ini}}{N} \sum_{i=0}^N
\exp \left\{ \int_0^t \rmd t' \, 
V_{\mathrm{T}} \left(n^{(i)}_{t'} \right) \right\} \delta_{n, n^{(i)}_t},
\label{eq_ex_Taylor_sampling}
\end{eqnarray}
where 
\begin{eqnarray}
V_{\mathrm{T}}(n) = - \gamma n 
- \frac{\sigma^2}{2} n(n-1)
+ \gamma (n-1)(n-2) 
+ \frac{\sigma^2}{2} (n-1),
\end{eqnarray}
and $\delta_{A,B}$ is the Kronecker delta function;
$w_\mathrm{ini}$ is determined by the initial condition, which will be explained later.
This is because the second term in the r.h.s. in \eref{eq_ex_Taylor_master_2} 
does not change the state $n$; this term plays a role only as a weighting factor.
Such splitting of the time-evolution equation has been discussed in \cite{Ohkubo2013},
and the term is called the Feynman-Kac term \cite{Liggett_book}.

As for the initial condition,  $\widetilde{p}(x,t=0) = x^m$ should be satisfied,
and then 
the initial particle number should be set as $m$ in the Monte Carlo simulation,
and we take
\begin{eqnarray}
w_\mathrm{ini} = m!.
\end{eqnarray}

In short, it is enough to make sample paths 
using the birth-death process in \eref{eq_ex_Taylor_birth_death}
and to evaluate \eref{eq_ex_Taylor_sampling};
the coefficients $\{ P_{\mathrm{T}}(n,t) \}$ in \eref{eq_ex_Taylor_expansion}
are calculated numerically via the Monte Carlo simulations.

Note that the birth-death process in \eref{eq_ex_Taylor_birth_death}
cannot be interpreted as a chemical reaction system.
It has been known that the Doi-Peliti formalism, 
which has been used to derive the dual process in \cite{Ohkubo2013},
is deeply related to stochastic processes for chemical reactions \cite{Tauber2005};
in this sense, the viewpoint of the basis expansion gives
a new dual process \eref{eq_ex_Taylor_expansion} naturally,
and it would be difficult to find the dual process via the Doi-Peliti formalism.
In addition, because of the factorial $n!$ in \eref{eq_ex_Taylor_expansion},
$P_{\mathrm{T}}(n,t)$ becomes considerably large compared with $P(n,t)$ 
in the conventional dual process \eref{eq_naive_master_equation};
although this is different from importance sampling methods,
this could be useful to perform the Monte Carlo simulations
because $P(n,t)$ takes very small value in general for large $n$,
and rare event sampling usually needs some numerical techniques.

\subsection{Hermite-type basis functions}

As a final example, a dual stochastic process based on Hermite polynomials is derived.
That is, the following basis expansion is employed here:
\begin{eqnarray}
\phi_n(x) = H_n(x),
\end{eqnarray}
where $\{H_n(x)\}$ is the Hermite polynomials
obtained from
\begin{eqnarray}
H_n(x) = (-1)^n \rme^{x^2} \frac{\rmd^n}{\rmd x^n} \rme^{-x^2}.
\end{eqnarray}
That is,
\begin{eqnarray}
\widetilde{p}(x,t) = \sum_{n=0}^\infty P_{\mathrm{H}}(n,t) H_n(x).
\label{eq_ex_Hermite}
\end{eqnarray}

The Hermite polynomials satisfy the following three-term recurrence relation \cite{Koekoek_book}:
\begin{eqnarray}
H_{n+1}(x) = 2x H_n(x) - 2n H_{n-1}(x),
\end{eqnarray}
and hence, we have
\begin{eqnarray}
x H_n(x) = \frac{1}{2} H_{n+1}(x) + n H_{n-1}(x).
\label{eq_ex_Hermite_recurrence}
\end{eqnarray}
In addition, the derivative of $H(x)$ w.r.t. $x$ gives \cite{Koekoek_book}
\begin{eqnarray}
\frac{\rmd}{\rmd x} H_n(x) = 2n H_n(x) - H_{n+1}(x).
\label{eq_ex_Hermite_derivative}
\end{eqnarray}
Using \eref{eq_ex_Hermite_recurrence} and \eref{eq_ex_Hermite_derivative},
the time-evolution equation for $\{P_{\mathrm{H}}(n,t)\}$ is obtained as
\begin{eqnarray}
\fl
\frac{\partial}{\partial t} P_{\mathrm{H}}(n,t) 
=& \frac{1}{2} \gamma (n-1) P_{\mathrm{H}}(n-1,t) 
+ \left( -\gamma n - \frac{\sigma^2}{2} (n-1)n\right) P_{\mathrm{H}}(n,t) \nonumber \\
\fl
&+ \left( \gamma (2n+1)(n+1) + \sigma^2 n(n+1)\right) P_{\mathrm{H}}(n+1,t) \nonumber \\
\fl
&+ \left( - 2 \gamma (n+1)(n+2) - \sigma^2 (2n+1)(n+1)(n+2)\right) P_{\mathrm{H}}(n+2,t) \nonumber \\
\fl
&+ 2(\gamma+\sigma^2)(n+1)(n+2)(n+3) P_{\mathrm{H}}(n+3,t) \nonumber\\
\fl
&- 2 \sigma^2 (n+1)(n+2)(n+3)(n+4) P_{\mathrm{H}}(n+4,t).
\label{eq_ex_Hermite_master_pre}
\end{eqnarray}

Note that the time-evolution equation in \eref{eq_ex_Hermite_master_pre}
does not satisfy the probability conservation law.
In addition, it is not enough to employ the technique used in the Taylor case
in order to recover the probabilistic characteristics;
for example, the final term in \eref{eq_ex_Hermite_master_pre}
changes the state $n \to n+4$,
but the sign of the term is minus 
and hence it should correspond to a negative transition rate.

The negative transition rate problem has also appeared in the Doi-Peliti formalism \cite{Ohkubo2013},
and it is possible to avoid the problem as below.

A new particle $n_0$ is added to the system,
and the following rules are employed:
\begin{enumerate}
\item Change the sign of the term related to the state change when we have the negative transition rate.
(Note that when the state is not changed, the term can be 
considered as a weighting (Feynman-Kac) term,
and there is no need to change the sign.)
\item Change the number of particle $n_0$ (add one particle)
for the terms which had the negative transition rate.
\end{enumerate}
In order to understand the above rule,
it would be easy to compare the following equation with \eref{eq_ex_Hermite_master_pre};
\begin{eqnarray}
\fl
\frac{\partial}{\partial t} P_{\mathrm{H}}(n,n_0,t) 
=& \frac{1}{2} \gamma (n-1) P_{\mathrm{H}}(n-1,n_0,t)
+ \left( -\gamma n - \frac{\sigma^2}{2} (n-1)n\right) P_{\mathrm{H}}(n,n_0,t) \nonumber \\
\fl
&+ \left( \gamma (2n+1)(n+1) + \sigma^2 n(n+1)\right) P_{\mathrm{H}}(n+1,n_0,t) \nonumber \\
\fl
&+ \left( 2 \gamma (n+1)(n+2) + \sigma^2 (2n+1)(n+1)(n+2)\right) P_{\mathrm{H}}(n+2,n_0-1,t) \nonumber \\
\fl
&+ 2(\gamma+\sigma^2)(n+1)(n+2)(n+3) P_{\mathrm{H}}(n+3,n_0,t) \nonumber \\
\fl
&+ 2 \sigma^2 (n+1)(n+2)(n+3)(n+4) P_{\mathrm{H}}(n+4,n_0-1,t)
\end{eqnarray}
Then, using the same discussion with the Taylor case,
we have the following birth-death process after some calculations:
\begin{eqnarray}
\begin{array}{ll}
n, n_0 \to n+1, n_0 & \textrm{at rate} \,\, \gamma n /2, \\
n, n_0 \to n-1, n_0 & \textrm{at rate} \,\, \gamma (2n-1)n + \sigma^2 (n-1) n,\\
n, n_0 \to n-2, n_0+1 & \textrm{at rate} \,\, 2\gamma (n-1)n + \sigma^2 (2n-3)(n-1)n, \\
n, n_0 \to n-3, n_0 & \textrm{at rate} \,\, 2(\gamma+\sigma^2)(n-2)(n-1)n, \\
n, n_0 \to n-4, n_0+1 & \textrm{at rate} \,\, 2 \sigma^2 (n-3)(n-2)(n-1)n
\end{array}
\label{eq_ex_Hermite_birth_death}
\end{eqnarray}
and 
\begin{eqnarray}
V_{\mathrm{H}}(n) =& - \frac{1}{2} \gamma n 
- \frac{\sigma^2}{2} n(n-1)
+ \gamma (2n-1)n \nonumber\\
&+ \sigma^2 (n-1)n 
+ 2\gamma (n-1)n + \sigma^2 (2n-3)(n-1)n \nonumber\\
&+ 2(\gamma+\sigma^2)(n-2)(n-1)n 
+ 2\sigma^2 (n-3)(n-2)(n-1)n.
\end{eqnarray}

In short, making $N$ sample paths from \eref{eq_ex_Hermite_birth_death},
the coefficients in \eref{eq_ex_Hermite} are obtained by
\begin{eqnarray}
P_{\mathrm{H}}(n,t) = 
\frac{1}{N} \sum_{i=0}^N
w_\mathrm{ini}^{(i)}\exp \left\{ \int_0^t \rmd t' \, 
V_{\mathrm{H}} \left(n^{(i)}_{t'} \right) \right\} \delta_{n, n^{(i)}_t} (-1)^{n^{(i)}_{0,t}},
\label{eq_ex_Hermite_sampling}
\end{eqnarray}
where $n^{(i)}_t$ and $n^{(i)}_{0,t}$ correspond to 
the particle number of $n$ and $n_0$ of $i$-th path at time $t$, respectively. 
The initial weights $\{w_\mathrm{ini}^{(i)}\}$ for $i$-th path should be chosen carefully, as denoted later.
The factor $(-1)^{n_{0,t}^{(i)}}$ means that 
the state change, which is caused by the negative transition rate, gives 
a contribution of a factor $(-1)$ 
and hence the virtual particle number $n_0$ plays a role as the change of the sign;
note that $n_{0,t=0}^{(i)} = 0$ at the initial time.
In addition, the initial condition should be chosen 
so that
\begin{eqnarray}
\sum_{n=0}^\infty P_{\mathrm{H}}(n,t=0) H_n(x) = x^m,
\end{eqnarray}
and, as discussed in the Taylor case,
the Monte Carlo results should be adequately scaled via $\{w_\mathrm{ini}^{(i)}\}$.
For example, when $m=1$,
we start from $1$ particle system in the Monte Carlo simulations,
and $w_\mathrm{ini}^{(i)} = 1/2$ for all $i$ because $H_1(x) = 2x$.
In order to evaluate the second order moment, the initial condition
and the weights $\{w_\mathrm{ini}^{(i)}\}$ become a little complicated.
Since $H_0(x) = 1$ and $H_2(x) = 4x^2 - 2$,
we have $x^2 = \frac{1}{4} H_2(x) + \frac{1}{2} H_0(x)$.
Hence, we could employ the following initial settings:
\begin{itemize}
\item Choose the initial particle number as $n_{t=0}^{(i)} = 0$ or $n_{t=0}^{(i)} = 2$ randomly
(i.e., with probability $1/2$ respectively).
\item If we choose $n_{t=0}^{(i)} = 0$, 
set $w_\mathrm{ini}^{(i)} = \frac{1}{2} \times \left(\frac{1}{2}\right)^{-1}$;
if $n_{t=0}^{(i)} = 2$, 
set $w_\mathrm{ini}^{(i)} = \frac{1}{4} \times \left(\frac{1}{2}\right)^{-1}$.
Note that the factor $\left(\frac{1}{2}\right)^{-1}$ is needed
in order to include the probability for the initial choice.
\end{itemize}

\section{Concluding remarks}

In the present paper, we give a new viewpoint for the duality
between stochastic differential equations and birth-death processes.
The viewpoint is based on the basis expansion,
which enables us to obtain various dual stochastic processes naturally.
As far as we know, the birth-death processes 
in \eref{eq_ex_Taylor_birth_death} and \eref{eq_ex_Hermite_birth_death}
have not been derived yet as the dual stochastic processes for \eref{eq_sFKPP}.
Additionally, the derivation is rather simple compared with the previous one \cite{Ohkubo2013};
the knowledge of the creation and annihilation operators are not needed,
and only the elemental calculus and the integration by parts are enough
to obtain the time-evolution operator for the dual process.
It was also shown that the negative transition rate problem is adequately avoided
using the similar technique in the previous work \cite{Ohkubo2013}.

In statistical physics, the duality relation has been used for calculating
physical quantities such as moments;
in some cases, it is possible to obtain analytical solutions 
for dual birth-death processes,
and the analytical solutions enable us to give the time-dependent or stationary solutions
for the original processes.
Note that the merit of the dual processes is not restricted
to the cases with analytically solvable systems.
We can see the important point in \eref{eq_basic_duality};
once the time-evolution of the dual birth-death process is evaluated,
the time-evolution of the original stochastic differential equation
\textit{with arbitrary initial conditions} is obtained.
As explained in section~2.3, this merit is used
to construct the nonlinear Kalman filter \cite{Ohkubo2015}.
In this sense, the numerical solutions for a little complicated
dual processes are enough for practical purposes.
In addition, we reformulated the time-evolution equation
so as to be suitable for the Monte Carlo simulations,
which will avoid the curse of dimensionality for multivariate cases.
As shown in the present paper, 
the non-uniqueness of the dual stochastic process is explicitly clarified.
It would be beneficial to find useful basis expansions
for each original stochastic process in future.
We expect that these extensions will open up a way
to seek practical and numerical studies 
of the usage of the duality concepts in stochastic processes.

\ack
This work was supported by JST, PRESTO Grant Number JPMJPR18M4, Japan.\\

\end{document}